\documentclass[10pt, conference]{IEEEtran}

\usepackage{subfig}
\usepackage{listings}
\usepackage{amsmath}
\usepackage{algorithmic}
\usepackage{times}
\usepackage{graphicx, color}
\usepackage{booktabs}
\usepackage{multirow}

\usepackage{ifthen,amssymb,color}
\newboolean{showcomments}
\setboolean{showcomments}{true}
\ifthenelse{\boolean{showcomments}}
  {\newcommand{\nb}[2]{
  \fbox{\bfseries\sffamily\scriptsize#1}
    {\sf\small$\blacktriangleright$\textit{\textcolor{red}{#2}}$\blacktriangleleft$}
   }
  }
  {\newcommand{\nb}[2]{}
   
  }

\title{Obfuscating Java Programs by Translating Selected Portions of Bytecode to Native Libraries}

\author{	\IEEEauthorblockN{		Davide Pizzolotto\IEEEauthorrefmark{1}\IEEEauthorrefmark{2},
		Mariano Ceccato\IEEEauthorrefmark{1}\\
		~
	}
	\IEEEauthorblockA{		\IEEEauthorrefmark{1}		Fondazione Bruno Kessler, Trento, Italy   \\
			}
	\IEEEauthorblockA{		\IEEEauthorrefmark{2}		University of Trento, Trento, Italy \\
			}

}

\begin{document}

\maketitle

\begin{abstract}
Code obfuscation is a popular approach to turn program comprehension and analysis harder, with the aim of mitigating threats related to malicious reverse engineering and code tampering. However, programming languages that compile to high level bytecode (e.g., Java) can be obfuscated only to a limited extent. In fact, high level bytecode still contains high level relevant information that an attacker might exploit.

In order to enable more resilient obfuscations, part of these programs might be implemented with programming languages (e.g., C) that compile to low level machine-dependent code. In fact, machine code contains and leaks less high level information and it enables more resilient obfuscations.

In this paper, we present an approach to automatically translate critical sections of high level Java bytecode to C code, so that more effective obfuscations can be resorted to. Moreover, a developer can still work with a single programming language, i.e., Java.

 \end{abstract}


\IEEEpeerreviewmaketitle

\section{Introduction}
\label{sec:intro}

Man-at-the-end (MATE~\cite{collberg-falcarin}) attacks are those threats to software security that occur when a program runs in a device controlled by the end user, potentially involving the user as attacker. These threats include malicious reverse engineering, for instance aiming at violating intellectual code property, and program tampering, to alter the code execution or the data used.

Code obfuscation is a widely adopted class of techniques to prevent or mitigate MATE attacks, by transforming the code to make it harder to understand and analyze, but without changing the code input-output behavior. However, very widely adopted programming languages compile to a high level bytecode, that is more difficult to obfuscate, because it contains detailed structural and type information. 

In particular, to meet the requirement of allowing runtime-time verification and runtime-time optimizations, Java bytecode has been designed so that it contains significant high level information. In fact, Java bytecode includes method and class identifiers, and variables are strongly typed. Moreover, class, fields and methods are referenced by name in clear text, rather than by address. Optimized machine code, obtained when compiling C code contains less information~\cite{vinciguerra2003experimentation}, because it is not meant to support runtime-time verification and run-time optimization. 

As a workaround to these disadvantages coming with Java bytecode, developers are often suggested to hide critical parts of a program by implementing them with other languages, that compile to lower level machine code, such as C\footnote{Android authority, How to hide your API Key in Android. https://www.androidauthority.com/how-to-hide-your-api-key-in-android-600583/}. However, this practice brings multiple drawbacks:
\begin{itemize}
\item A developer has to master an additional programming language, different than the mainstream language used in the rest of the software project;
\item When portions of the program are written in C and linked as an external library, many compile-time and load-time verifications normally performed by Java are impossible. This exposes the program to potential errors, whose checking is usually delegated to the Java compiler;
\item C code is usually not portable, often it is written for and runs on a single architecture, or a sub-set of architectures. Writing portable C code would need additional effort, that may increase development cost.
\end{itemize}

In this paper, we present an approach to automatically translate portions of Java bytecode to C. Depending on the peculiar confidentiality constraints of a program, a developer is supposed to annotate those segments of a Java software project that are required to be translated to C. After compiling the Java code, our tool detects the annotated parts and removes them from the bytecode. These parts are automatically translated to C and converted to a shared library, that is referenced by the rest of the Java program using the JNI interface. 

Differently from other approaches (e.g., Caffeine~\cite{hsieh1996java}) that aim at translating a whole Java program to C code, we only translate {\em portions} of the program and we preserve the compatibility between the Java and C parts. This allows the remaining Java part to meet framework or interface requirements potentially imposed by the execution environment. For instance, Android requires apps to extend system classes (e.g., in {\em Activities}) and some Java methods to adopt specific signatures.

This translation represents a first step towards the mitigation of malicious reverse engineering, because more effective obfuscations can apply on translated C code, for instance using publicly available obfuscation tools for C~\cite{Banescu2016Obfuscation,Viticchie2016Reactive,tiella2017opaque,sutter2016reference}.

On the other hand, the full automation of our translation overcomes the disadvantages of manually writing C code in Java projects:
\begin{itemize}
\item Developers only write and maintain Java code. They do not have to care about C code, because it is automatically generated based on the Java one;
\item The translation is done at the bytecode level, after the Java code is compiled. So the Java compiler is still able to perform all the compile-time verifications, and notify errors and warnings to the developer;
\item A requirement of our translation is to generate fully portable C code, that can be easily compiled for many distinct platforms.
\end{itemize}

Working at bytecode level brings and additional advantage. Our approach applies not only to programs developed in Java, but also written in the programming languages that compile to Java bytecode, e.g., Kotlin, Scala, JRuby, Clojure, and others.

Our paper is structured as follows. Section~\ref{sec:background} covers the technical background of the adopted technology. Section~\ref{sec:approach} presents our approach and Section~\ref{sec:results} its empirical validation. After presenting related work in Section~\ref{sec:related}, Section~\ref{sec:conclusion} closes the paper.

 \section{Background}
\label{sec:background}

Java code is not compiled to architecture-specific machine code (like C and C++ do) but to an intermediate representation called bytecode. Java bytecode instructions are then interpreted and executed by the Java Virtual Machine, which is available for each specific execution platform. In this section, we cover some background on the Java language and on its runtime environment that is relevant for this paper.

\subsection{Stack Based}
\label{sec:stack}

The Java Virtual Machine~\cite{lindholm2014java} is stack based. It means that the data structure where the operands are stored is a stack. All the operations are performed by popping (consuming) data from the stack, processing them and pushing the results back onto the stack. For instance, to add two integer values, these values should be pushed to the stack, then the {\em integer-addition} opcode is called. Two values are popped from the stack and the addition result is pushed on top of it, where it is available for the next opcode.

Method calls use the stack similarly to arithmetic operations. Before calling a method, all the actual parameters should be already pushed to the stack. The method return value is available in the stack after the method execution is complete (in case the called method has a non-void return type).

\subsection{Opcodes}

The Java bytecode language supports a rich set of opcodes, that are summarized in the following.

{\bf Field access}: opcodes used to read/write a class field. The class name, field name and its type are explicit parameters of this opcode. The field value is read from (or written to) the stack.

{\bf Method call}: opcodes used to call a method. The called method name and the name of the class that defines the called method are parameters of the instruction. The formal parameter types are part of the method name. The actual parameters are expected to be available on the stack before the call, and the return value (if any) is on top of the stack when the called method returns. For non static methods, a pointer to the current class is also an actual parameter on the stack. The opcode for evaluating dynamic language methods, such as lambda expressions in the Java language, is a special case of method call.

{\bf New instance}: opcodes to allocate new object instances and new arrays. Once allocated, the pointer to an object/array is available in the stack.

{\bf Control flow}: opcodes related to the definition of labels and of conditioned/unconditioned jumps to labels. A specific opcode supports the switch-case Java instruction, by specifying the full jump table. Each value of the switch expression is assigned a specific label to jump to. Among these opcodes we have also the return opcodes.

{\bf Constants}: opcodes used to push a constant (string or numeric) value into the stack. Different opcodes are used for different constant types (String, int, float and so on).

{\bf Local variables}: opcodes to read and write local variables. Local variable are read/written by coping their values from/to the stack, or by loading constant values.

{\bf Expressions}: opcodes related to arithmetic (e.g., sum) and logical (e.g., xor) operations, type conversion (e.g., from long to int, or from long to float). A special expression is the \texttt{INSTANCEOF} opcode that checks whether a given object is of a given (dynamic) type.

{\bf Exception handling}: opcodes related exceptions. They are used to define the extent of the {\em try} block, the corresponding {\em catch} block and the type of the exception handled. Additionally, an opcode is used to throw a new exception, that the caller context might handle.

\subsection{Java Native Interface}

The Java Native Interface (or JNI for short) is a framework that allows Java code to execute {\em native} code, i.e. architecture-specific machine code, typically written in C. The aim of this framework is to let the JVM run portions of code that are platform-specific, for instance because they interact directly with the underlying hardware or because they do low-level I/O.

JNI is executed when invoking a Java method that has the {\em native} modifier. This method has empty method body and, whenever called, its corresponding platform-specific binary code will be executed, from a shared library. Figure~\ref{fig:jni-example} shows an example of native code. The Java code on the left-hand side defines two methods, {\em sub} and {\em add}, both of them accept two integer parameters and return an integer value. While method {\em sub} is fully implemented in Java, method {\em add} lacks a Java implementation. Since second method has the {\em native} modifier, a platform-dependent implementation is expected. 

Its platform-dependent implementation in C is shown in the right-hand side of Figure~\ref{fig:jni-example}. Whenever a Java {\em native} method is executed, the JNI framework delegates a C function following a strict naming convention. The C function name is obtained by concatenating the {\em ``Java\_''} prefix, with the Java class name and the Java native method name. In the example, the C function {\em Java\_Calculator\_add} corresponds to the method {\em add} in the class {\em Calculator}.

The C function formal parameter list includes a pointer to the Java execution environment ({\em *env}), a pointer to the object that is called ({\em thisObj}) and then the formal parameters of the Java method for which a machine-dependent implementation is provided ({\em a} and {\em b}).

The reference {\em *env} to Java execution environment can be used to interact with the JVM, for instance to inspect structure of classes, to access their fields and to call their methods. Moreover, utility functions are available in the environment to probe execution status, for instance to know if an exception has been thrown, and to perform type conversion. 

Each Java native type is converted to a C type of the same size, available to the C developer as a macro, e.g. {\em int} variables are visible as {\em jint}, signed 32-bit integer. Objects are instead visible as type {\em jobject}.

The C code in the example, just performs the sum in C and then it returns the numeric value to the Java calling context.

\begin{figure}[t]
	\centering
	\begin{tabular}{cc}
		\begin{lstlisting}[linewidth=.5\linewidth,language=java,basicstyle=\scriptsize\ttfamily,breaklines=false,escapeinside={(*@}{@*)}]
class Calculator{

  int sub(int a, int b){
    return a - b;
  }

  native int add(int a, int b);	
  
}
\end{lstlisting} 		&
		\begin{lstlisting}[linewidth=.4\linewidth,language=c,basicstyle=\scriptsize\ttfamily,breaklines=false,escapeinside={(*@}{@*)}]
#include <jni.h>

JNIEXPORT jint JNICALL 
Java_Calculator_add(
        JNIEnv *env, 
        jobject thisObj, 
        jint a, jint b) {
    jint c = a + b;
    return c;
}
\end{lstlisting} \\
	(a) Java code &
	(b) C code \\
			\end{tabular}
	\caption{Example of integration of C code in Java with JNI.}
	\label{fig:jni-example}
\end{figure}

 \section{Transformation}
\label{sec:approach}

This section presents our approach to transform a Java program by removing selected parts of it, and by replacing them with a native C library.

\subsection{Overview}

\begin{figure}[htb]
  \centering
  \includegraphics[width=0.6\columnwidth]{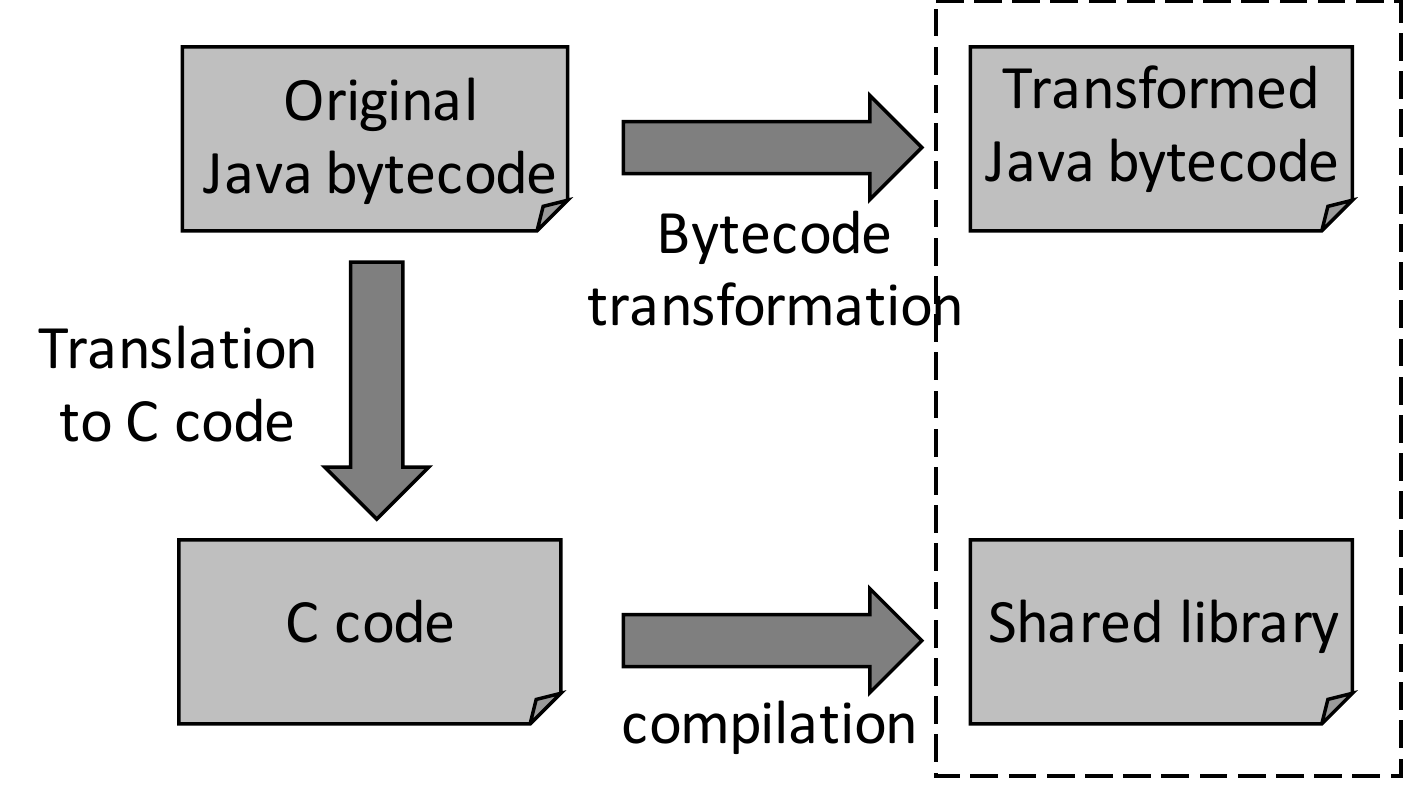}
  \caption{Overview of the transformation approach.}
  \label{fig:overview}
\end{figure}

\begin{figure*}[htb]
  \centering
  \begin{tabular}{cccc}
    \begin{lstlisting}[linewidth=.5\linewidth,language=Java,basicstyle=\scriptsize\ttfamily,breaklines=false,escapeinside={(*@}{@*)}]
class Calculator {

  @Obfuscate
  int sum(int x, int y){
    z=x+y;
    return z;
  }

}
\end{lstlisting}
  &
    \begin{lstlisting}[linewidth=.5\linewidth,language=JVMIS,basicstyle=\scriptsize\ttfamily,breaklines=false,escapeinside={(*@}{@*)}]
@Obfuscate
sum(II)I
  ILOAD 1
  ILOAD 2
  IADD
  ISTORE 3
  ILOAD 3
  IRETURN
\end{lstlisting}
 & 
    \begin{lstlisting}[linewidth=.5\linewidth,language=JVMIS,basicstyle=\scriptsize\ttfamily,breaklines=false,escapeinside={(*@}{@*)}]
native sum(II)I

static <clinit>
  LDC "libSim.so"
  INVOKESTATIC 
    java/lang/System.loadLibrary
    (Ljava/lang/String;)V
  RETURN
\end{lstlisting}
 & 
    \begin{lstlisting}[linewidth=.5\linewidth,language=JVMIS,basicstyle=\scriptsize\ttfamily,breaklines=false,escapeinside={(*@}{@*)}]
#include <jni.h>

JNIEXPORT jint JNICALL Java_Calculator_add(
        JNIEnv *env, jobject thisObj,
        jint x, jint y) {
  jvalue vars[4]
  Push(vars[1]);
  Push(vars[2]);
  Push( (int)Pop() + (int)Pop());
  vars[3] = Pop();
  Push(vars[3]);
  return (int)Pop();
}
\end{lstlisting}
 \\
    (a) Java source code & 
    (b) Java bytecode&
    (c) Transformed Java bytecode&
    (d) Translated C source code\\
  \end{tabular}
  \caption{Example of transformation of Java bytecode and the corresponding translated C code.}
  \label{fig:example}
\end{figure*}

The overview of our approach is shown in Figure~\ref{fig:overview} and shown in an example of Figure~\ref{fig:example}. We assume that those Java methods requiring obfuscations are annotated by developers with the annotation {\em @Obfuscate} (method {\em sum} in Figure~\ref{fig:example}(a)). Our transformation tool reads the compiled Java bytecode (Figure~\ref{fig:example}(b)) and finds these annotations. 

The annotated methods are rewritten, by removing their body and by adding the modifier {\em native} so that JNI will be activated at runtime, as shown in Figure~\ref{fig:example}(c). Moreover, a call to {\em System.loadLibrary} is added in the class static initialization section (i.e., the {\em \textless clinit\textgreater} method), to bind this class with the native C library that the translator will create.

The removed Java body is then translated to a C function, opcode by opcode, and a C source file is emitted as shown in Figure~\ref{fig:example}(d). The C code is compiled to a shared library, and this library is loaded as part of the static initialization of the rewritten Java class. 

The final transformed program consists of the dashed area in Figure~\ref{fig:overview}, i.e. the rewritten Java bytecode and the compiled C library.

\subsection{Java Operand Stack Emulation}
\label{sec:stack-emulation}

Our transformation is implemented as an iteration on the opcodes from the body of the method to translate and applies opcode-by-opcode. For each opcode found in the Java bytecode, a set of C statements are appended to the new C function. Since the Java Virtual Machine relies on the stack (see Section~\ref{sec:stack}), also the C code generated by the translator works on a stack, that emulates in C the original Java operand stack. Thus, opcodes in the original Java bytecode that push/pop data are translated as corresponding push/pop operations on the C stack.

While the Java operands stack is used to store operands and results, Java local variables are stored in an array and accessed by index with the \texttt{LOAD} and \texttt{STORE} opcodes. Similarly to the operands stack, also the local variables array is replicated by the C code generated by the translator. 

The size of the operands stack and of the variables array is known at compile time and thus they are implemented in our C layer as statically allocated arrays.
We just introduced a small difference, due to the different size of variables between the Java bytecode and the JNI environment exposed to C code. In Java bytecode, the operands stack and the variables array contains elements of 32-bit. Java values that are 64-bit long ({\em double} and {\em long}) consume two elements. 

However, the JNI support for numeric types is different, because all the Java types are represented in JNI as a C union of 64-bit. For this reason, our operands stack and our variable array are defined in C as array of 64-bit elements, and any value consumes just one element.

The example in Figure~\ref{fig:example} shows how we translate a portion of Java that involves the operand stack. In the right-hand side, the original Java source code adds two local variables {\em x} and {\em y} and assigns the result to {\em z}, then this value is returned. This compiles to the Java bytecode in Figure~\ref{fig:example}(b), \texttt{ILOAD} opcodes load integer variables 1 and 2 (corresponding to {\em x} and {\em y}) in the operands stack. These operands are consumed by the integer addition \texttt{IADD} and the result is pushed to the stack. This value is read by \texttt{ISTORE} and written to the variable 3 (corresponding to {\em z}). Then \texttt{IRETURN} reads the computed integer value from the stack and returns it. 
The C source code generated by our translator is shown in Figure~\ref{fig:example}(d). \texttt{ILOAD} and \texttt{ISTORE} opcodes are translated to pushes and pops to copy values from the operands stack to local variables {\em vars} and vice-versa. The \texttt{IADD} opcode is translated as the integer C sum, with 2 pops from the stack (operands) and 1 push to the stack (result), similarly to what the original \texttt{IADD} operand did in Java. The \texttt{IRETURN} opcode is translated into a C {\em return}.

\subsection{Arithmetic}

In Java, every arithmetic operation is compiled to a bytecode opcode of a specific type (e.g, integer, float and so on). Each bytecode opcode can be directly translated to an arithmetic operator available in C. This simple translation is possible, because the set of arithmetic operations in the Java bytecode and operators in the C language are the same. For instance, in Figure~\ref{fig:example} the integer sum in Java is translated with the integer sum in C.

The only special cases are logical and arithmetic shifts. While Java bytecode supports both kind of shifts, C applies one or the other depending on the operand type being signed or unsigned. In our tool, this has been solved with a type cast, to be consistent with the Java semantics.

\subsection{Control Flow}

Control flow manipulation includes opcodes providing conditional branching and looping.
While Java provides keywords to implement a structured control flow (e.g., {\em while} and {\em for}), when Java code is compiled to bytecode, only \emph{label}s, \emph{goto}s and \emph{if} branches are left. C supports goto and labels, so these opcodes are directly translated to the corresponding C keywords.

Java bytecode supports two switch opcodes, namely \texttt{TABLESWITCH} and \texttt{LOOKUPSWITCH}. 
They are translated with canonical C {\em switch} statements. Although the Java language supports {\em switch} statements with String values, that in C are not supported, they are compiled to bytecode switches on integer values, so the same translation pattern applies.

It is also worth noting that unnecessary labels must not be translated, when they appear in dead Java bytecode (e.g., after a return statement), because this is not allowed in C.

\subsection{Method Call and Field Access}

When the annotated code calls other Java methods or accesses fields of Java classes a special translation is required, because methods and fields are directly visible to Java code, but they are not directly visible to C code. The workaround is to call methods and access fields by name using reflection features available in the JNI utility functions.

\begin{figure*}[htb]
  \centering
  \begin{tabular}{cccc}
    \begin{lstlisting}[linewidth=.5\linewidth,language=Java,basicstyle=\scriptsize\ttfamily,breaklines=false,escapeinside={(*@}{@*)}]
int x = sum(1, 2);
return x;
\end{lstlisting}
  &
    \begin{lstlisting}[linewidth=.5\linewidth,language=JVMIS,basicstyle=\scriptsize\ttfamily,breaklines=false,escapeinside={(*@}{@*)}]
ALOAD 0
ICONST_1
ICONST_2
INVOKEVIRTUAL B.sum (II)I
ISTORE 1
ILOAD 1
IRETURN
\end{lstlisting}
 & 
    \begin{lstlisting}[linewidth=.5\linewidth,language=JVMIS,basicstyle=\scriptsize\ttfamily,breaklines=false,escapeinside={(*@}{@*)}]
Push(var[0]);
Push(1);
Push(2)
jvalue par[2];
par[1] = Pop();
par[0] = Pop();
jvalue target = Pop();
CallIntMethodA("MyClass", "sum", "(II)I", target, par);
var[1] = Pop();
Push(var[1])
return (int)Pop();
\end{lstlisting}
 \\
    (a) Java source code & 
    (b) Java bytecode&
    (c) C source code\\
  \end{tabular}
  \caption{Example of translation of method call.}
  \label{fig:call}
\end{figure*}

Figure~\ref{fig:call} shows the Java source code and bytecode used for calling method {\em sum}, and the C code resulting from our translation. Java bytecode uses the \texttt{INVOKE*} opcodes to perform method calls. When this opcode is interpreted and executed by the JVM, the number of parameters required by the called methods (two parameters in the example) are popped automatically from the operands stack and the result of the call, if any, is pushed back onto the stack. 

In C, instead, the function call expects actual parameters as an array. So, the array {\em par} is created before calling the function, and two {\em Pop}s instructions are added to fill the parameter array. An additional {\em pop} instruction is added to retrieve the {\em target}, i.e. the class instance that contains the method to call. Then, the Java method can be called using the JNI utilities with the parameter array and the target object. After the call is returned, the return value is explicitly pushed to the stack.

Some invocation variants are available in Java bytecode, so they require a special translation.

{\bf Virtual calls}: 
Virtual method calls are those for which class inheritance matters. This is the case of opcodes \texttt{INVOKEVIRTUAL}, \texttt{INVOKESTATIC} and \texttt{INVOKEINTERFACE}. In this cases, the inheritance tree of the called class should be visited, and the most specific method implementation should be called. For this purpose, the JNI layer provides the utility functions {\em Call[type]MethodA}, where {\em [type]} should be replaced the return type of the called Java method (primitive type, Object or Void). The only difference between \texttt{INVOKEVIRTUAL} and \texttt{INVOKESTATIC} is that the first requires an instance of the called class, while the second does not because it calls a static Java method. \texttt{INVOKEINTERFACE} works the same as \texttt{INVOKEVIRTUAL}.

{\bf Non-virtual calls}: 
Non-virtual method calls are calls to a specific method belonging to a class, without descending inheritance tree. They are used for example to call parent class' methods.
Non-virtual method calls are supported by the \texttt{INVOKESPECIAL} opcode. The translation is similar to the case virtual calls. A first difference is that a different JNI utility function is used, i.e. {\em CallNonvirtual[type]MethodA}. 

A special case is when \texttt{INVOKESPECIAL} is used to call a class constructor, because, memory should be allocated for the new object to be created. In the Java bytecode, before calling a constructor, space is allocated with the opcode \texttt{NEW}. While in Java bytecode allocation and constructor call are two separate actions, in the JNI a single utility function {\em NewObject} encloses both of them.

{\bf Field Access}:
Field access is translated using reflection, similarly to method calls, using JNI utility functions {\em Set[type]Field}, {\em SetStatic[type]Field}, {\em Get[type]Field} or {\em GetStatic[type]Field}. Different functions are available to read or write instance fields and static fields, and different functions are available for fields of different types.

{\bf Arrays}:
Java arrays are allocated on the heap, similarly to Java objects. The allocation of one-dimensional arrays with the opcodes \texttt{NEWARRAY} and \texttt{ANEWARRAY} can be translated to a call the JNI function \emph{New[type]Array}, with the array size as parameter. 
The opcode \texttt{MULTIANEWARRAY}, used to allocate a multidimensional array, is translated as multiple calls to the aforementioned function. The \texttt{ARRAYLENGTH} opcode to read array length is translated to the JNI function \emph{GetArrayLength}.

\subsection{Exception Detection}

In Java, exceptions implicitly alter the control flow, because whenever an exception is thrown, the execution is interrupted and transferred to an exception handling block. However, Java exceptions do not interrupt the control flow in the C code. So, the translation has to explicitly simulate the Java model to preserve original control flow and the original program semantic.

To translate the Java exception model in C, we need to distinguish three cases that can occur:
\begin{itemize} 
\item {\em Thrown exceptions}: exceptions created and thrown in the annotated Java method (to be translated) by the \texttt{ATHROW} opcode;
\item {\em Forwarded exceptions}: exceptions that are created and thrown by a method that is called by the translated method, so they occur after method calls. In Figure~\ref{fig:jni-exception}(a), {\em MyException} is an example of forwarded exception, thrown by the called method; and 
\item {\em System exceptions}: exceptions that can happen after a statement that is not a method call, such as division by zero after arithmetic operations. In Figure~\ref{fig:division-by-zero}(a), {\em ArithmeticException} is an example of system exception, thrown by an expression statement.
\end{itemize}

The first problem is to detect that an exception is thrown. For each case, a different translation strategy apply.

{\bf Thrown exceptions}:
These exception are easily detected statically, when the \texttt{ATHROW} opcode is found.

{\bf Forwarded exceptions}:
Forwarded exceptions are created by a called Java method and they are not handled by it, so they are forwarded to the caller, i.e. the method translated to C. 

Since the invoked method runs on the JVM layer, in order to detect that one of these exceptions occurred, we have to use the JNI utility function {\em ExceptionOccurred} after each method call. 

{\bf System exceptions}: 
System exceptions can be thrown potentially by non-call opcodes. Since checking for exceptions after every instruction would be computationally too expensive, these checks must be minimized.

Since all system exceptions are subclasses of \emph{RuntimeException}, a first optimization consists in only checking for this exception type in C code, and not for all the possible sub-types. By manually inspecting the Java documentation\footnote{Java SE 8 RuntimeException subclasses list and documentation. https://docs.oracle.com/javase/8/docs/api/java/lang/RuntimeException.html}, we listed all the opcodes after which we should check for an exception. They are for example array access for \emph{IndexOutOfBoundsException}, divisions for \emph{ArithmeticException} or \texttt{CHECKCAST} opcodes for \emph{ClassCastException}. The second optimization consists in checking for system exceptions only after this subset of opcodes.

\subsection{Try-Catch Blocks}

Figure~\ref{fig:jni-exception}(a) shows an example of Java code with a {\em try-catch} block. The {\em try} block contains the exception handled code. In case it throws an exception, the corresponding {\em catch} exception handling code is executed. In Java {\em catch} blocks are typed, in the example the type is {\em MyException}.

Figure~\ref{fig:jni-exception}(b) shows the compiled Java bytecode. The {\em try-catch} block is compiled to a \texttt{TRYCATCH} opcode, with three labels and a type as arguments. \texttt{L0} and \texttt{L1} are labels that delimit the {\em try} block, \texttt{L2} is the label of the {\em catch} block and \texttt{MyException} is the type of the caught exception.

{\bf Thrown exceptions}:
An \texttt{ATHROW} opcode that creates a new exception is translated simply by creating an exception object in C. Then, the type of the exception object is inspected using the {\em IsInstanceOf} JNI utility function, in order to jump to the appropriate exception handler block. In case no block is found for this exception type, the JVM is informed about the exception with a {\em ThrowNew} JNI function and the execution of the transformed method is interrupted. The Java calling context is supposed to handle this exception.

{\bf Forwarded exceptions}:
Figure~\ref{fig:jni-exception} shows an example of how forwarded exceptions are translated. Forwarded exceptions are translated similarly to thrown exceptions, but with a different detection method. When the exception is caught by the C code, the JVM must be informed about that with a call to the JNI function {\em ExceptionClear}. Otherwise, the exception is propagated back to the Java calling method.

Additionally, when an exception is caught by the C code, the operands stack must be cleared and the JVM exception added with the JNI function {\em ExceptionDescribe}, in order to keep the JVM stack and translated stack aligned.

In the example of Figure~\ref{fig:jni-exception}, the value returned by {\em InvokeStatic} is not the output of the function, which is pushed onto the operands stack, but an integer which is non-zero if an exception occurred. 

Then, if this value is non-zero, the {\em AlignWithJVM} creates a copy of the exception for the JNI layer, and the exception type is compared with every catch block with the {\em InstanceOf(stack, ``MyException'')} call.
If there is a match, the JVM is informed about the fact that the exception is handled locally with a call to {\em ClearException} and the control jumps to the catch block. Otherwise, if no catch block matches this exception type, the function returns and the caller Java code will handle this exception.

\begin{figure}[htb]
  \begin{tabular}{c}
    \begin{lstlisting}[linewidth=.5\linewidth,language=Java,basicstyle=\scriptsize\ttfamily,breaklines=false,escapeinside={(*@}{@*)}]
try {
  MyClass.MethodThrowingException();
} catch (MyException e) {
  return 1;
}
return 0;
\end{lstlisting}
 \\
    (a) Java code \\
    \\
    \begin{lstlisting}[linewidth=.5\linewidth,language=JVMIS,basicstyle=\scriptsize\ttfamily,breaklines=false,escapeinside={(*@}{@*)}]
TRYCATCHBLOCK L0 L1 L2 MyException
L0:
INVOKESTATIC MyClass.methodThrowingException ()V
L1:
ICONST_0
IRETURN
L2:
ICONST_1
IRETURN
\end{lstlisting}
 \\
    (b) Java bytecode\\
    \\
    \begin{lstlisting}[linewidth=.5\linewidth,language=JVMIS,basicstyle=\scriptsize\ttfamily,breaklines=false,escapeinside={(*@}{@*)}]
int exception;
jvalue params0[0];
exception = InvokeStatic("MyClass", 
  "methodThrowingException", "()V", params0);
if (exception) {
  AlignWithJVM(stack);
  if (InstanceOf(stack, "MyException")) {
    ClearException();
    goto L2;
  } 
  else {
    //here the exception is not cleared
    //caller Java code will handle this exception
    return;
  }
}
Push(0);
return (int)Pop();
L2:
Push(1);
return (int)Pop();
\end{lstlisting}
 \\
    (c) Translated C code\\
  \end{tabular}
  \caption{Forwarded exception in Java code, Java bytecode and its translation in C.}
  \label{fig:jni-exception}
\end{figure}

{\bf System exceptions}:
Inheritance tree of system exception is known at transformation time. In fact, system exceptions are defined by the JVM implementation and not by user code, so we do not need to inspect their type by dynamically calling {\em IsInstanceOf}. Thus, we can solve every {\em catch} block directly at transformation time. 

In our support C functions that could throw a system exceptions (such as {\em IDiv} to compute division between integer values), a zero return value means correct execution, while a return value greater than zero means that an exception occurred.

In case of non-zero return value, the exception inheritance tree is considered at transformation time. If exception happened inside a {\em try} block, a {\em goto} is added to the corresponding {\em catch} block (whose type is known at transformation time). Otherwise, an exception is created and this method returns, so the exception is passed to the Java calling context. 

An example of this translation can be seen in Figure~\ref{fig:division-by-zero}. In the example, the {\em IDiv} function returns non-zero if an exception occurred, and this is verified at runtime in a conditional branch. The label to jump to is decided in the following way: since the candidate exception is {\em ArithmeticException}, we search in the code for catch blocks for all the corresponding supertypes, they are {\em ArithmeticException}, {\em RuntimeException}, {\em Exception}, {\em Throwable}. The translator found a catch block for {\em ArithmeticException} at label {\em L2}, so the branch is a {\em goto} to the matching catch block, i.e. {\em goto L2}. 

\begin{figure*}[htb]
  \begin{tabular}{ccc}
    \begin{lstlisting}[linewidth=.5\linewidth,language=Java,basicstyle=\scriptsize\ttfamily,breaklines=false,escapeinside={(*@}{@*)}]
int retval;
try {
  retval = x/y;
} catch (ArithmeticException e) {
  retval = 0;
}
\end{lstlisting}
     &
  \begin{lstlisting}[linewidth=.5\linewidth,language=JVMIS,basicstyle=\scriptsize\ttfamily,breaklines=false,escapeinside={(*@}{@*)}]
TRYCATCHBLOCK L0 L1 L2 ArithmeticException
L0:
ILOAD 1
ILOAD 2
IDIV
L1:
IRETURN
L2:
ICONST_0
IRETURN
\end{lstlisting}
   &
  \begin{lstlisting}[linewidth=.5\linewidth,language=JVMIS,basicstyle=\scriptsize\ttfamily,breaklines=false,escapeinside={(*@}{@*)}]
Push(vars[1]);
Push(vars[2]);
int exception = IDiv(stack);
if(exception)
  goto L2;
return Pop();
L2:
Push(0);
return Pop();
\end{lstlisting}
   \\
    (a) Java code & (b) Java bytecode & (c) Translated C code
  \end{tabular}
  \caption{System exception in Java code, Java bytecode and its translation in C.}
  \label{fig:division-by-zero}
\end{figure*}

 \section{Empirical Validation}
\label{sec:results}

In this section, we demonstrate the feasibility of our approach by adopting it on some software projects, and by collecting some performance measurements. We assess our approach by investigating the following research questions:

\begin{itemize}
\item {\bf RQ$_{arch}$}: Do translated code execute correctly on different architectures?
\item {\bf RQ$_{time}$}: What is the performance overhead of a transformed program?
\item {\bf RQ$_{scale}$}: Does performance scale when translating a larger and larger part of a program?
\end{itemize}

The first search question {\bf RQ$_{arch}$} is meant to check that translation does not introduce errors in the programs. This should be checked on different architectures to confirm that the portability requirement of Java is still met on translated code.

Translation is expected to come with the cost of some performance overhead. {\bf RQ$_{time}$} is meant to investigate how longer programs take to execute, after they are transformed. 

The last research question {\bf RQ$_{scale}$} measures how the performance degrades when a larger and larger portion of a program is subject to translation. This dimension is important to tune an optimal trade-off between adequate obscurity and acceptable performance degradation.

Other properties of translated/obfuscated code, such us how harder it is to understand and to attack, are out of the scope of the present paper. To investigate these properties, human studies and controlled experiment are required, and we are planning and conducting them as part of our research activity~\cite{ceccato2014family,ceccatoSCAM2016,icpc2017}

\subsection{Case Studies}

In our experiments, we consider three open source Java applications. Even though our approach applies to compiled Java bytecode, we decided to opt for open source projects because they come with test cases. We rely on test cases to assess execution correctness of translated code. 

Table~\ref{tab:apps} shows some data about our case studies. For each case study (first column), the table shows the number of classes (second column), the total number of lines of Java code (third column) and the number of available test cases (last column).

\begin{table}[ht]
\centering
\begin{tabular}{crrr}
  \toprule
Name & Classes & LoCs & Tests \\ 
  \midrule
  Joda Time & 247 & 15,016 & 4,222 \\ 
  Java2word &  53 & 1,165 & 308 \\ 
  JFreeChart & 658 & 52,041 & 2,176 \\ 
   \bottomrule
\end{tabular}
\caption{Case studies.} 
\label{tab:apps}
\end{table}
 
The case study programs are:

\begin{itemize}
  \item {\em Joda Time}: A date and time library using several calendar systems, used to be the de-facto standard prior to Java SE 8 and its \emph{java.time};
\item {\em Java2word}: A library used to create Word documents from Java code and export its XML representation; and
\item {\em JFreeChart}: A library used to create and display charts in Java applications, written with either Swing or JavaFX.
\end{itemize}

\subsection{Experimental Setting}

When defining our experimental setting, we aimed at identifying meaningful transformation configurations, effective to answer our research questions. In particular, we meant to translate only methods that can be tested to a large extent.

To achieve this objective, we ran the full test suite of each case study with Emma\footnote{EMMA: a free Java code coverage tool. http://emma.sourceforge.net/} and we collected coverage data. Only those methods whose code coverage was at least 70\% were considered as candidates for translation. We also collected code data such as (i) the method length in terms of number of basic blocks and LoCs and (ii) the number of times each method is executed by test cases.

Annotated methods are translated from Java to C, according to the annotation syntax accepted by our tool, as described in Section~\ref{sec:approach}. In our experiment, we considered different translation configurations, corresponding to different sets of annotated methods. We developed a small annotation tool, that injects annotations directly into Java bytecode, according to the experimental configurations decided based on test coverage, method length and number of executions.

As soon as the Java bytecode is translated to C (and compiled), all the test cases are executed again to make sure that translation did not alter the program correctness.

\subsection{Correctness of Translated Code}

Even if test first was adopted, and a large number of unit tests have been used to check translation at development time, we mean to validate correctness also on real world case studies.

The translated C code should execute correctly on multiple architecture, as the original Java program did. In particular, this includes processors with different architectures and instruction sets, with 32 bit or 64 bit registers length. Thus, we considered the following test hosts:
\begin{itemize}
\item {\bf x86\_64}: MacBook Pro with an Intel i5-4258U @ 2.40GHz CPU, running a 64 bit operating system;
\item {\bf ARM}: Raspberry Pi 1 model B with a Broadcom BCM2835 SoC mounting a 500Mhz ARMv6 CPU (overclocked to 700Mhz), running Raspbian version 32 bit.
\end{itemize}

All the test cases pass for all the three case studies on every test host, so we can answer RQ$_{arch}$ as follows:
\begin{center}
\fbox{
\begin{minipage}[t]{0.9\linewidth}
\textit{Translated C code executes correctly on processors with different instruction sets, with 32 bit and 64 bit architectures.}
\end{minipage}
}
\end{center}

\subsection{Performance Overhead of Transformed Code}

To measure performance overhead, we defined an transformation configuration for each case study, consisting of a set of methods with the following requirements. First of all, we consider only methods that are tested quite exhaustively, so we require that at least 70\% of their code is covered by test cases. These methods are sorted by code length in ascending order and the top larger 10 methods are selected. These 10 methods are annotated to be translated.

Then, we ran the original code and the transformed code on the two reference hosts and we collected the test execution time. The measurement has been repeated 100 times to reduce random errors and increase the accuracy.

Table~\ref{tab:time.arch} shows the result of this experiment with a distinct line per each case study on each host. The table shows the case study name and the host on the first and second columns respectively. Then, the table reports the average time (in seconds) and standard deviation for original code (third and fourth columns) and for the translated code (fifth and sixth columns). The last column reports the relative time increase due to transformation.

\begin{table}[ht]
\centering
\begin{tabular}{cc|rr|rr|r}
  \toprule
\multirow{ 2}{*}{Name} & \multirow{ 2}{*}{Arch} 
& \multicolumn{2}{c|}{Original} & \multicolumn{2}{c|}{Transformed} & \multirow{ 2}{*}{Delta} \\
&      & mean & sd & mean & sd &  \\ 
  \midrule
  Joda Time & x86\_64 & 0.82 & 0.03 & 0.96 & 0.04 & +18\% \\ 
  Joda Time & ARM & 20.55 & 0.30 & 24.84 & 0.31 & +21\% \\ 
  Java2word & x86\_64 & 0.68 & 0.02 & 0.77 & 0.02 & +14\% \\ 
  Java2word & ARM & 8.26 & 0.26 & 9.09 & 0.18 & +10\% \\ 
  JFreeChart & x86\_64 & 3.85 & 0.42 & 4.03 & 0.47 & +5\% \\ 
  JFreeChart & ARM & 40.34 & 0.64 & 42.33 & 0.58 & +5\% \\ 
   \bottomrule
\end{tabular}
\caption{Execution time for original and transformed programs.} 
\label{tab:time.arch}
\end{table}
 
As we can observe, the case study with larger time increase is {\em Joda Time} with an average performance overhead between +18\% and +21\%, on x86\_64 and on ARM respectively. {\em Java2word} has a lower performance overhead, between +10\% and +14\%. A minor performance overhead of +5\% has been observed for {\em JFreeChart}, and in x86\_64 host the time difference between transformed and original code is lower than the standard deviation.

Considering these results, we can answer RQ$_{time}$ as follows:
\begin{center}
\fbox{
\begin{minipage}[t]{0.9\linewidth}
\textit{The performance overhead due to code transformed are larger (+[18-21]\%) for {\em Joda Time}, lower for {\em Java2word} (+[10-14]\%) and minor for {\em JFreeChart} (+5\%).}
\end{minipage}
}
\end{center}

\subsection{Relation between Translation Extent and Speed}

The last research question investigates the relation between the ratio of the program subject to translation and the execution-time overhead. To study this phenomenon, we conceived a number of different configurations with an increasing number of annotated methods. As in the previous experiment, we only consider methods with code coverage above 70\%. However, instead sorting them by size, we now sort methods by the number of times they are executed by the test suites. In fact, we are interested in studying the worst case scenario, when the translated code is crucial for the program because it is executed a lot of times and might largely impact the user experience.

For this experiment we only consider one of the three case studies, i.e. {\em Joda Time}, the one with most significant performance overhead recorded so far. The first configuration is the original code. The next configuration is one with just one translated method, i.e. the method that is the most executed by the test suite. Then, a new configuration is defined by translating another method, and so on until we have in total 64 configurations. Each configuration has been executed 100 times.

\begin{figure}[htb]
\begin{center}
\includegraphics[width=\columnwidth,trim={0 0.6cm 0 2cm},clip]{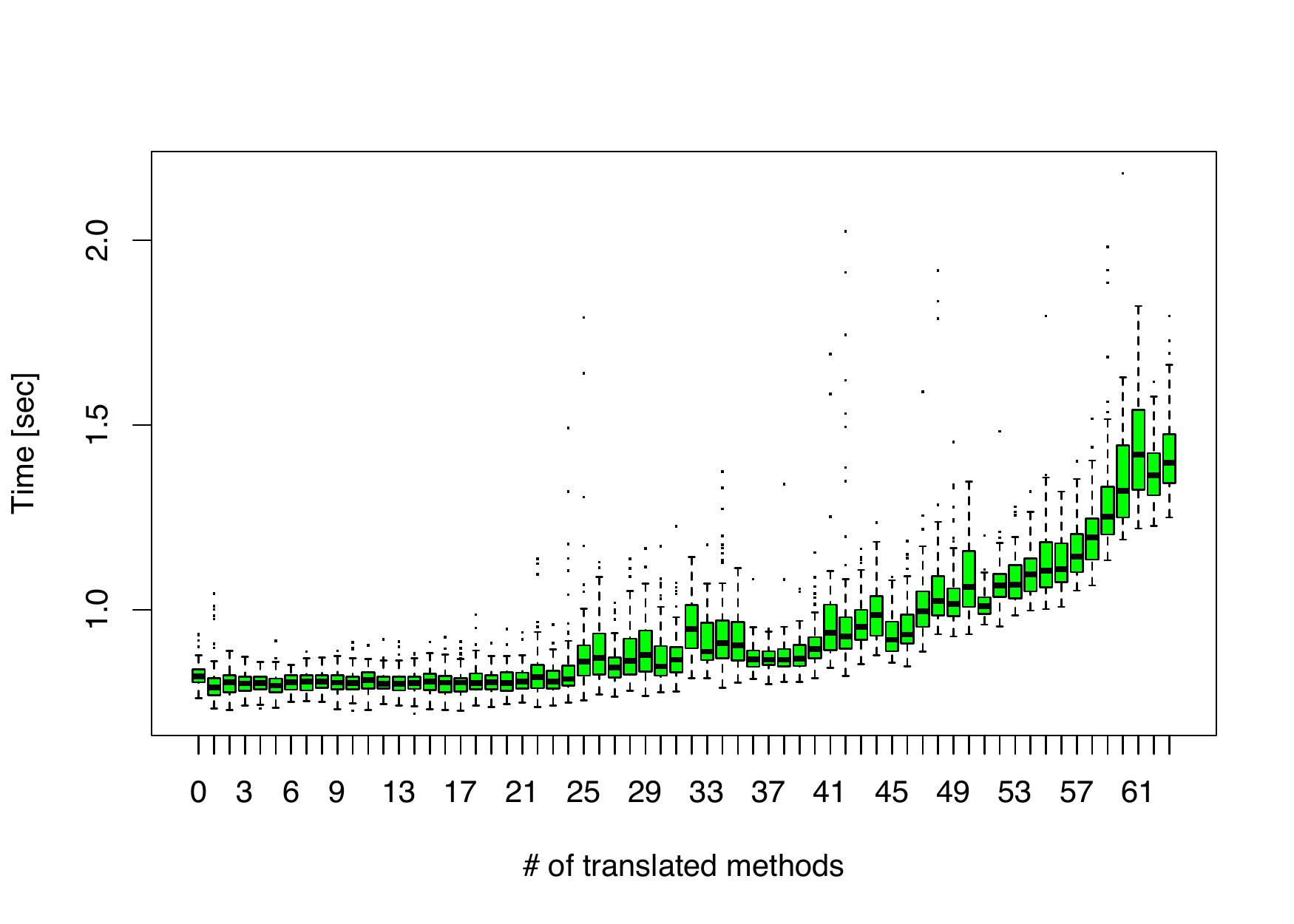}

\end{center}
\caption{Test suite execution time with increasing number of translated methods.}
\label{fig:boxplot-exec-time}
\end{figure}

The box plot in Figure~\ref{fig:boxplot-exec-time} reports the time taken to execute the full test suite. As we can see, for the first 20 configurations there is not major performance overhead because transformed program takes similar time to run as the original program (first configuration at the left-hand side). Then execution time starts to increase significantly with the number of translated methods.

\begin{figure}[htb]
\begin{center}
\includegraphics[width=\columnwidth,trim={0 0.5cm 0 2cm},clip]{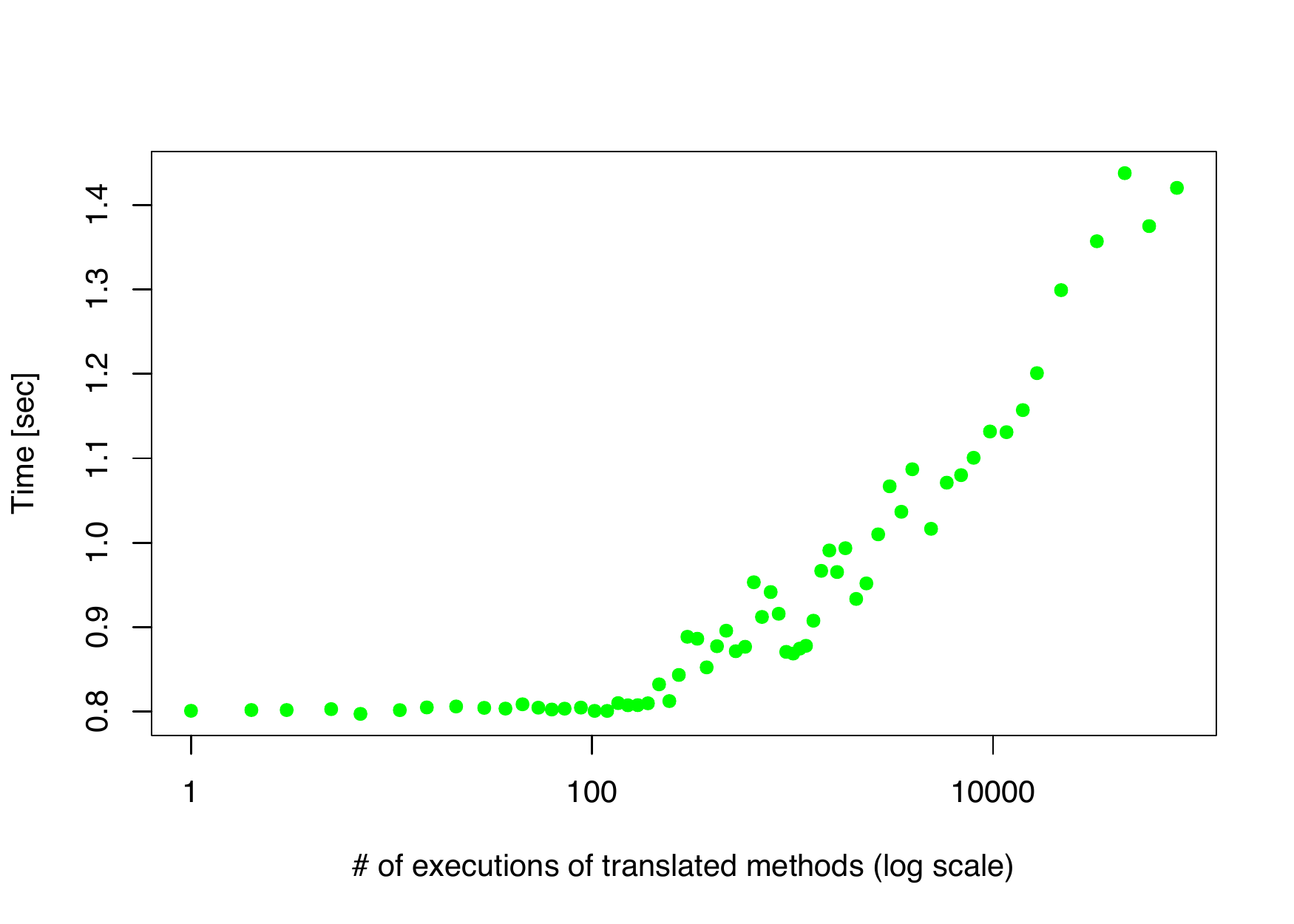} 
\end{center}
\caption{Test suite execution time with increasing number of executed translated methods.}
\label{fig:ex-vs-time}
\end{figure}

The same data can also observed from a different perspective. In particular, we are interested to study how many times the translated methods have been executed. Figure~\ref{fig:ex-vs-time} shows this trend. The x-axis reports (in logarithmic scale) the number of time that translated methods are executed. The y-scale reports the execution time of the test suite (in seconds). As we can see, below the 100 executions of translated methods, no noticeable overhead is evident. Then, an increasing trend can be observed.

\begin{figure}[htb]
\begin{center}
\includegraphics[width=\columnwidth,trim={0 0.5cm 0 2cm},clip]{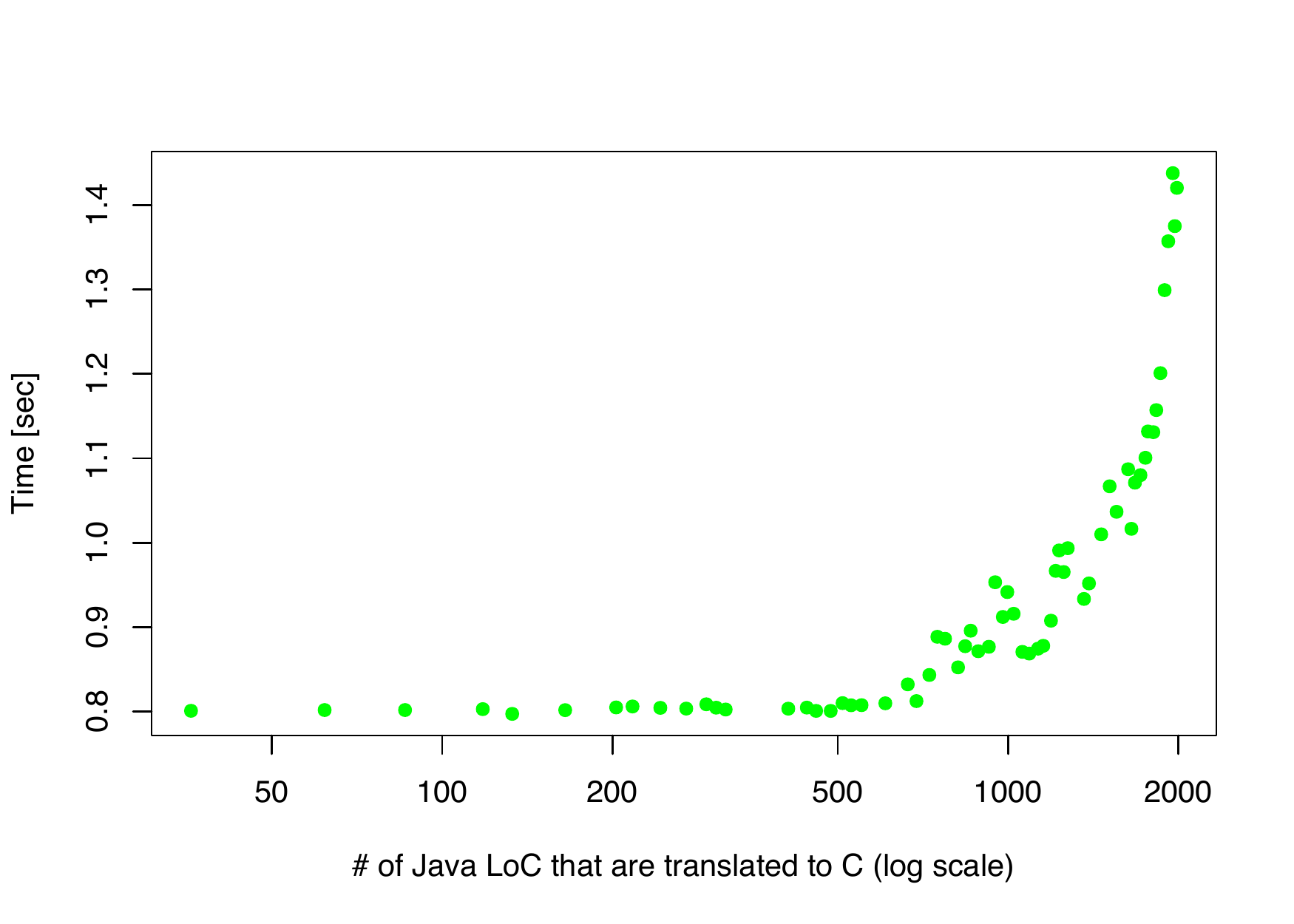} 
\end{center}
\caption{Test suite execution time with increasing number of Java lines that are translated.}
\label{fig:loc-vs-time}
\end{figure}

The last perspective is shown in Figure~\ref{fig:loc-vs-time}, where the considered configurations are shown in terms of the lines of Java code that are subject to code translation, still with the x-axis in logarithmic scale. Performance overhead is minor when less than 1000 LoCs are translated. Then an increase is observed.

Considering all these results, we can answer RQ$_{scale}$ as follows:
\begin{center}
\fbox{
\begin{minipage}[t]{0.9\linewidth}
\textit{Performance overhead due to code translation is negligible when the portion of translated code is limited. In {\em Joda Time} we observed negligible runtime overhead for less than 20 methods that, in total, are executed less than 100 times and that consist in approximately 1000 LoCs.}
\end{minipage}
}
\end{center}

 \section{Related Work}
\label{sec:related}

Obfuscation is used to make application code obscure so that it is more complex to
understand by a potential attacker who wants to reverse engineer it. Obfuscation techniques change  code structure without changing its functional behavior through different kinds of code transformations~\cite{wang2001protection,valdez1999software}. It is well-known that for binaries that mix code and data, disassembly and de-compilation are undecidable in the worst case~\cite{linn2003obfuscation}. On the other hand, under specific and restrictive conditions, some work reported that de-obfuscation is an NP-easy problem~\cite{appel2002deobfuscation}. Further, it was proven that a large number of functions cannot be obfuscated~\cite{barak2001pop}.

Many algorithms for code obfuscation have been proposed in the literature. In the taxonomy by Collberg et al.~\cite{collberg1997taxonomy}, they have been classified into layout, data and control-flow obfuscation.

The most related category is {\em layout obfuscation}. This category includes transformations that change or remove useful information from the intermediate language code or the source code without affecting the instructions that contribute to the actual computation. Usually removing debugging information, comments, and scrambling/renaming identifiers fall within the domain of layout obfuscation. 

Identifier renaming~\cite{chan2004advanced} is an instance of layout obfuscation that removes relevant information from the code by changing the names of classes, fields and operations into meaningless identifiers, so as to make it harder for an attacker to guess the functionalities implemented by different parts of the application. 

{\em Data obfuscation} category of transforms targets data and data structures contained in the program. Using these transformations, data encoding can be changed~\cite{tiella2017opaque}, variables can be split or merged, and arrays can be split, folded, and merged.

The objective of {\em control-flow obfuscation} is to alter the flow of control within the code. Reordering statements, methods, loops and hiding the actual control flow behind irrelevant conditional statements classify as control-flow obfuscation transforms.  Obfuscation based on Opaque predicates~\cite{Collberg1998} is a control-flow obfuscation that tries to hide the original behavior of an application by complicating the control flow with artificial branches. An opaque predicate is a conditional expression whose value is known to the obfuscator, but is hard to deduce statically by an attacker.  An opaque predicated can be used in the condition of a newly generated {\em if} statement. One branch of the {\em if} statement contains the original application code, while the other contains a bogus version of it. Only the former branch will be executed, causing the semantics of the application to remain the same. In order to generate resilient opaque predicates, pointer aliasing can be used, since inter-procedural static alias analysis is known to be intractable.

With the increasing adoption of Java as a programming language, the idea of translating Java bytecode to C was investigated~\cite{hsieh1996java}, but mainly as a way avoid the overhead due to the Java interpreter, i.e. by turning the whole Java program into a single machine-dependent executable. In our approach, we instead keep the original Java program structure, and only selected portions are translated to C to turn analysis more difficult and to enable more efficient obfuscations.

Other obfuscation approaches that rely on code translation are based on an obfuscated Virtual Machine~\cite{anckaert2006proteus,hu2006secure} (OVM for short). Binary machine-dependent code is translated to a custom opcodes that can be interpreted by the OVM. (A portion of) the clear code is replaced by the corresponding custom opcodes and the OVM is appended to the program. When the obfuscated program is launched, the OVM takes control: it reads, decodes and executes the custom opcodes. Such OVM can make it much harder to reverse-engineer programs because standard disassemblers and standard tracing tools (e.g., debuggers) do not target the custom opcodes, and because the attackers are not familiar with them.

These types of VMs are susceptible to OVM replacement attacks, in which an attacker replaces the original OVM that implements a number of security features by one that lacks those features~\cite{ghosh2012replacement}.

This attack is possible, because the application itself is not dependent to the specific OVM. As a countermeasure, techniques have been proposed to inject such bindings~\cite{ghosh2013software}.

Furthermore, these OVMs are susceptible to tracing attacks~\cite{raber2013virtual,sharif2009rotalume}, in which attackers collect and analyze execution traces to separate VM engine code (e.g., the code that maintains the software cache) from the original application code being executed in the software cache. From the remaining application code trace, they can then reconstruct the original program.

 \section{Conclusion}
\label{sec:conclusion}

Obfuscation is a quite popular technique to protect programs from malicious reverse engineering. However, programming languages such as Java that compile to highly structured bytecode still leak a lot of information to a potential attacker even if delivered in obfuscated form.

In this paper, we proposed an approach based on program transformation to translate Java bytecode to C. Essentially, security sensitive portion of code (annotated as such by the developer) are automatically removed from the Java program and translated to C. This C code is then compiled to machine-dependent binary code, and executed whenever the original code is called. 

As future work, we intend to pipeline this transformation tool with available obfuscation tools for C, to complete the tool-chain and enable very strong obfuscations on Java programs. Moreover, we plan to design and conduct controlled experiments and user studies to measure how much harder is our code to reverse engineer and attack than the original Java code.  
 
\section*{Acknowledgement}
This work has partially been supported by the activity ``API Assistant'' of the action line Digital Infrastructure of the EIT Digital and the GAUSS national research project, which has been funded by the MIUR under the PRIN 2015 program (Contract 2015KWREMX).

\bibliographystyle{abbrv}
\bibliography{refs}

\end{document}